\documentstyle[12pt]{article} 
\newcommand{\nn}{\nonumber}
\newcommand{\bd}{\begin{document}}
\newcommand{\ed}{\end{document}}
\newcommand{\bc}{\begin{center}} 
\newcommand{\ec}{\end{center}}
\newcommand{\be}{\begin{eqnarray}}
\newcommand{\ee}{\end{eqnarray}}
\newcommand{\eqn}{\global\def\theequation}
\newcommand{\sw}{sin^2 \theta_W}
\newcommand{\fbd}{f_B}
\renewcommand{\thefootnote}{\alph{footnote}}
\newcommand{\se}{\section}
\newcommand{\sse}{\subsection}
\newcommand{\bi}{\bibitem}
\def\kpnn{K\to\pi\nu\bar{\nu}}
\def\knn{K_L\to\pi^0\nu\bar{\nu}}
\def\kplus{K^+\to \pi^+\nu\bar{\nu}}
\def\klznu{K_L\to\pi^0 z^0\to\pi^0\nu\bar{\nu}}
\def\kznn{K\to\pi Z^0\to\pi\nu\bar{\nu}}
\def\kpznu{K^+\to\pi^{+}Z^0\to\pi^+\nu\bar{\nu}}
\def\knng{K_L\to\gamma\nu\bar{\nu}}
\begin{document}
\tolerance=10000
\baselineskip=7mm
\begin{titlepage}  

 \vskip 0.5in   
 \null
\begin{center}
 \vspace{.15in}
{\LARGE {\bf $K_L\to\gamma\nu\bar{\nu}$ in the Light Front Model
}
}\\
\vspace{1.0cm}
  \par
 \vskip 2.5em
 {\large
  \begin{tabular}[t]{c}
{\bf C.~Q.~Geng, C.~C.~Lih, and C.~C.~Liu
}
\\
\\
{\sl Department of Physics}
\\
{\sl National Tsing Hua University} 
\\  {\sl   Hsinchu, Taiwan, Republic of China }\\
\\
   \end{tabular}}
 \par \vskip 5.0em
 {\Large\bf Abstract}
\end{center}

We study the CP conserving and violating contributions to the decay of
$\knng$ in the standard model. In our analysis, we use 
the form factors for $K\to\gamma$ transitions calculated directly in
the entire physical range of momentum transfer
within the light front model. We find that
the branching ratios for the CP conserving and violating parts
are about $1.0\times 10^{-13}$ and $1.5\times 10^{-15}$, respectively.

\end{titlepage}

\se{Introduction}
$\ \ \ $


With the prospect of a new generation of ongoing kaon experiments,
a number of rare kaon decays have been suggested to test the
Cabibbo-Kobayashi-Maskawa (CKM) \cite{ckm} paradigm.
However it is sometimes a hard task to extract
the short-distance contribution, which depends on the CKM matrix, because
of large theoretical uncertainties in the long-distance contribution to
the decays \cite{review}.
To avoid this difficulty, much of recent theoretical work as well as
experimental attention has been on searching for
the two modes: $\kplus$ and $\knn$. 
It is believed that the long-distance contributions in these two modes are 
much smaller than the short-distance ones and therefore they are 
negligible \cite{ld,wise,ghl,ghl2}.

It has been shown that the decay branching ratio of $\kplus$
is close to $10^{-10}$  \cite{bg,others_k}
arising dominated from the short-distance loop contributions containing 
virtual charm and top quarks. This decay is a CP conserving process and 
probably the cleanest one, in the sense of theoretical uncertainties,
to study the absolute value of the CKM element $V_{td}$.
Currently,
the E787 group at BNL \cite{E787}
has seen one event for the decay with the branching ratio of
$B(\kplus)= 1.5\begin{array}{c}+3.5\\-1.3\end{array}\times 10^{-10}$, 
which is consistent with the
standard model prediction, and it is expected that there will be several
events when the analysis is complete. 
The approved experiments of E949 at BNL and E905 at FNAL \cite{BCP-K} 
will have the sensitivities of $10^{-11}$ and $10^{-12}$, respectively.

On the other hand, the decay $\knn$ depending on the imaginary part of 
$V_{td}$ is a CP violating process \cite{knn}
and offers a clear information about the origin
of CP violation. In the standard model, it is dominated by the Z-penguin
and W-box loop diagrams with virtual top quark and the
decay branching ratio is found to be at the level of $10^{-12}$
\cite{bg,others_k},
whereas the current experimental limit is less than $5.9\times 10^{-7}$
given by the experiment of E799 at FNAL \cite{E799-K}.
Several dedicated experimental searches \cite{BCP-knn} for this decay mode
are underway at KEK, BNL and FNAL, respectively.
However, from an experimental point of view very challenging
efforts are necessary to perform the experiments.
This is because all the final state particles are neutral and
the only detectable particles are $2\gamma$'s from $\pi^0$.

As an alternative search, it was proposed \cite{geng-1} 
to use the decay of $K_L\to\pi^+\pi^-\nu\bar{\nu}$. But, the
decay branching ratio is small and  the background for
$\pi^+\pi^-$ is large \cite{Hsiung}.
In this paper, we study the radiative decay of $\knng$, where there is one
photon at the final states.
The mode has been considered previously in Refs. \cite{MO,RP,MP}
and it is believed that the decay is short distance dominated
\cite{RP,MP}. 
However, the decay branching ratio predicted in Ref.
\cite{RP} and \cite{MP} does not agree with each other.
Furthermore, all the discussions were confined in the CP conserving
contribution due to the vector part of the structure-dependent
amplitudes \cite{RP,MP}. The decay branching ratio was found at the
levels of $10^{-11}$ and $10^{-13}$ in Refs. \cite{RP} and \cite{MP},
respectively, which are about two orders of magnitude different.  
To clarify the issue, we will re-examine the decay by using
the form factors of $K\to\gamma$ transition calculated directly in
the entire physical range of momentum transfer
within the light front framework. 
We will study both CP conserving and violating contributions to
the decay branching ratio, respectively.

The paper is organized as follows. In Sec.~2,
we present the relevant effective Hamiltonian for the radiative
decay of $\knng$ and study
the form factors in the $K^0\to\gamma$ transition
within the light front framework.
In Sec.~3, we calculate the decay branching ratio. We also compare our
result with those in literature \cite{RP,MP}.
We give our conclusions in Sec.~4.


\se{Effective Hamiltonian and Form Factors}
The processes of $K_L\to\nu_l\bar{\nu}_l\gamma$ ($l=e\,,\mu\,,\tau$),
arise from the box and 
$Z$-penguin diagrams
that contribute to $s\to d\nu_l\bar{\nu}_l$ with the photon 
emitting from the charged particles in the diagrams. 
The effective Hamiltonian for $s\to d\, \nu\bar{\nu}$ at the quark
level in the standard model is given by 
\be
H_{eff}(s\to d\, \nu\bar{\nu}) &=&
{G_F\over \sqrt{2}}{\alpha\over   
2\pi\sin^2\theta_W}
\sum_{l=e,\mu,\tau}\left(
\lambda_c\;X_{NL}^l+\lambda_t\;X(x_t)\right)
\nn\\
&&\cdot\bar{d}\gamma_{\mu}(1-\gamma_5)s\bar{\nu}_l\gamma^{\mu}
(1-\gamma_5)\nu_l\,,
\label{he2}
\ee
where $x_t=m_t^2/M_W^2$, $\lambda_i=V_{is}^*V_{id}\ (i=c,t)$ represent
the products of the CKM matrix elements, and
the functions of $X_{NL}^l$ and $X(x_t)$ correspond to the top and charm
contributions in the loops with the next-to-leading logarithmic
approximation, respectively, and their expressions can be
found in Ref. \cite{Bu}. 
In the Wolfenstein parameterization, we have
\be
Re\lambda_c &=&-\lambda (1-{\lambda^2\over 2})\,,
\nn\\
Re\lambda_t&=&- (1-{\lambda^2\over 2})A^2\lambda^5(1-\rho+{\lambda^2\over 
2}\rho)\,,
\nn\\
Im\lambda_c&=&Im\lambda_t\;=\;A^2\lambda^5\eta\,.
\ee
For Phenomenological applications, we use
\be
X(x_t) &=& \eta_X\cdot X_0(x_t)
\ee
where
\be
\eta_X &=& 0.994\,,\nn\\
X_0(x_t)&=& {x_t\over 8}\left[-\,{2+x_t\over 1-x_t}+{3x_t-6\over
(1-x_t)^2}\ln 
x_t\right]\,,
\label{Dxt}
\ee
with the $\overline{MS}$ definition of the top-quark mass,
$m_t\equiv \bar{m}_t(m_t)=(166\pm5)\ GeV$.
For the charm sector, from the Table 1 in Ref. \cite{Bu},
for example, one has 
\be
X_{NL}^{e,\mu}&=& 11.00\times 10^{-4}\,,\nn\\
X_{NL}^{\tau}&=& 7.47 \times 10^{-4}\,,
\ee
with the central values of the QCD scale
$\Lambda=\Lambda^{(4)}_{\overline{MS}}=(325\pm80)\ MeV$ and the charm
quark mass $m_c=\bar{m}_c(m_c)=(1.30\pm0.05)\ GeV$.

 From the effective Hamiltonian in Eq. (\ref{he2}), we 
see that to find the decay rate, we have to evaluate the hadronic matrix 
element: $<\gamma|J_{\mu}|K^0>$, where 
$J_{\mu}=\bar{d}\gamma_{\mu}(1-\gamma_5)s$.
The element can be parameterized as follows:
\be
<\gamma (q)|\bar{d}\gamma^{\mu }\gamma_{5}s|K^0(p+q)>
&=& -e{F_{A}\over M_{K}}\left[
\epsilon ^{*\mu }( p\cdot q) -( \epsilon ^{*}\cdot p)
q^{\mu }\right]
\nn\\
<\gamma (q)| \bar{d}\gamma^{\mu}s|K^0(p+q)> &=&
-ie{F_{V}\over M_{K}} \epsilon^{\mu \alpha \beta \gamma }\epsilon 
_{\alpha }^{*}p_{\beta }q_{\gamma } 
\label{n2}
\ee
where $q$ and $p+q$ are photon and $K$-meson four 
momenta, $F_{A}$ and $F_{V}$ are form factors of axial-vector and vector,
respectively, and $\epsilon $ is the photon polarization vector. 

The form factors of $F_A$ and $F_V$ in Eq. (\ref{n2}) can be calculated in
the light front quark model at the time-like momentum transfers in which
the physically accessible kinematic region is 
$0\leq p^{2}\leq p_{\max}^{2}$
and they are found to be \cite{lih1,lih2,hsu} 
\be
	F_{A}(p^{2}) &=& -4M_{K}
		\int \frac{dx'd^{2}k_{\bot }}{2(2\pi)^{3}}\Phi
		\left( x,k_{\bot }^{2}\right) {1\over 1-x}
			\nonumber \\
	&&~~~~~~~~ \times \left\{ \frac{1}{3}\frac{-m_{s}+Bk_{\bot }^{2}
		\Theta}{m_{s}^{2}+
	k_{\bot}^{2}}-\frac{2}{3}\frac{m_{d}-Ak_{\bot }^{2}\Theta}
		{m_{d}^{2}+k_{\bot }^{2}}  \right\}\,, \label{fffa}
\ee
\be
	F_{V}(p^{2}) &=&4M_{K}
		\int \frac{dx'd^{2}k_{\bot }}{2\left( 2\pi \right) ^{3}}\Phi
		\left( x,k_{\bot }^{2}\right) {1\over 1-x}
			\nonumber \\
	&&\left\{ \frac{1}{3}\frac{-m_{s}-(1-x)(m_{s}-m_{d}) k_{\bot }^{2}
	\Theta }{m_{s}^{2}+k_{\bot }^{2}}-\frac{2}{3}\frac{m_{d}-
		x\left( m_{s}-m_{d}\right) k_{\bot }^{2}\Theta }{m_{d}^{2}
		+k_{\bot }^{2}}\right\}\,, \label{fffv}
\ee
where
\be
	A &=& (1-2x')x(m_s-m_d) -2x'm_d\,, \nn\\
	B &=& [(1-2x') x-1]m_s+(1-2x') (1-x)m_d\,,\nn\\
	\Phi (x,k_{\bot}^2) &=& N\left( 
		{2x(1-x) \over M_0^2-(m_d-m_s)^2}\right)^{1/2}
		\sqrt{{dk_{z}\over dx}}\exp \left( 
		-{\vec{k}^{2}\over 2\omega_K^2}\right)\,, \nn\\
	\Theta &=& {1\over \Phi(x,k_{\bot}^2) }
		{d\Phi(x,k_{\bot}^{2})\over dk_{\bot}^2} \, , \nn\\
	 x&=&x'\left(1-{p^2\over M_K^2}\right),\
		\vec{k}=(\vec{k}_{\bot},\vec{k}_{z}) \,,
\ee
with 
\be
        N &=& 4 \left({\pi\over \omega_K^{2}}\right)^{3\over 4}\,,
                \nn \\
        k_{z} &=&\left( x-\frac{1}{2}\right)
        M_{0}+\frac{m_{s}^{2}-m_{d}^{2}}{2M_{0}} \,, 
\nn \\
 M_0^2&=&{k^2_{\bot}+m_d^2\over x}+{k^2_{\bot}+m_s^2\over 1-x}\,,
\ee
and $\omega_K$ being chosen to be $0.34\ GeV$ fixed by the
decay constant of $f_K=160\ MeV$.

To illustrate the form factors, we input the values of $m_{d}=0.3$, 
$m_{s}=0.4$, and $M_{K}=0.5$ in $GeV$ to integral 
whole range of $p^{2}$.  It is interesting to note that 
at $p^2=0$, we get that $(F_A(0),F_V(0))=(0.0429,0.0915)$
comparing with $(0.0425,0.0945)$ found in the chiral
perturbation theory at the one-loop level \cite{Bi93}.

\se{Decay Branching Ratios}

 From the effective Hamiltonian for 
$K^0\to\gamma \nu\bar{\nu}$ in Eq. (\ref{he2}) and
the form factors defined in Eq. (\ref{n2}), 
we can write the amplitude of $K^0\to\gamma \nu\bar{\nu}$ as
\be
M(K^0\to \gamma \nu\bar{\nu}) &=&
i{G_F\over \sqrt{2}}{\alpha\over   
2\pi\sin^2\theta_W}
\sum_{l=e,\mu,\tau}\left(
\lambda_c\;X_{NL}^l+\lambda_t\;X(x_t)\right)
\nn\\
&& \cdot\epsilon_{\mu }^{*}H^{\mu \nu }\bar{u}(p_{\bar{\nu}}) 
\gamma_{\nu }(1-\gamma_{5})v(p_{\nu})\,,
\ee
with
\be
H_{\mu \nu } &=&{F_A\over M_K}(-p'\cdot q\,g_{\mu\nu}+p'_{\mu }q_{\nu })
+i\epsilon _{\mu \nu \alpha \beta }\frac{F_{V}}{M_K}q^{\alpha
}p^{'\beta}\,. 
\ee
where $p'$ is the four momentum of $K^0$ and 
the form factors $F_{A,V}$ are given by Eqs. (\ref{fffa}) and
(\ref{fffv}), respectively.
Since $K_L\simeq K_2=(K^0-\bar{K}^0)/\sqrt{2}$, we may write
\be
{\cal M}(K_L\to \gamma \nu\bar{\nu}) &=& {\cal M}_{CPC}+{\cal }M_{CPV}
\ee
where ${\cal M}_{CPC}$ and ${\cal M}_{CPV}$ are the amplitudes
corresponding to CP conserving and violating contributions, respectively,
which are given
by
\be
{\cal M}_{CPC}&=&-{G_F\over \sqrt{2}}{\alpha\over   
2\pi\sin^2\theta_W}
{2\over \sqrt{2}}\sum_{l=e,\mu,\tau}\left(
Re\lambda_c\;X_{NL}^l+Re\lambda_t\;X(x_t)\right)
\nn\\
&&\cdot\epsilon^{\mu\nu\alpha\beta}\frac{F_{V}}{M_K}
\epsilon_{\mu}^{*}q_{\alpha}p_{'\beta}\,
\bar{u}(p_{\bar{\nu}}) 
\gamma_{\nu }(1-\gamma_{5})v(p_{\nu})\,,
\ee
and
\be
{\cal M}_{CPV}&=&-{G_F\over \sqrt{2}}{\alpha\over   
2\pi\sin^2\theta_W}
{6\over \sqrt{2}}
Im\lambda_t\;X(x_t)
\nn\\
&&\cdot{F_A\over M_K}\epsilon_{\mu }^{*}
(-p'\cdot q\,g^{\mu\nu}+p'^{\mu }q^{\nu})
\bar{u}(p_{\bar{\nu}}) 
\gamma_{\nu }(1-\gamma_{5})v(p_{\nu})\,.
\ee
Here we have neglected the imaginary part of $Im\lambda_c$ for
${\cal M}_{CPV}$.

To evaluate the branching ratio, one needs to replace
$p^{2}$ into $(p',q)$.
In the physical allowed region of $\knng$,
one has that
\be
0\leq p^{2}\leq M_{K}^{2}\,.
\ee
In the $K_L$ rest frame, the partial decay rate of $\knng$
is given by
\be
d^2\Gamma =\frac{1}{(2\pi )^{3}}\frac{1}{8M_{K}}\mid {\cal M}\mid
^{2}dE_{\gamma}dE_{\nu }\,.
\label{n25}
\ee
where we have used two variables to describe the kinematic of the decay.
For convention, we define
$x_{\gamma}=2E_{\gamma}/M_K$ and
$x_{\nu}=2E_{\nu}/M_K$ 
as the normalized energies of the photon and and neutrino, respectively,
and we have the form
\be
p^{2}&=& M_K^{2}(1-x_{\gamma})\,.
\ee
The differential decay rate is then given by
\be
\frac{d^{2}\Gamma}{dx_{\gamma}dx_{\nu} }
&=&\frac{M_K}{256\pi^{3}}\left| {\cal M}\right| ^{2}\,.
\ee
By integrating the variable $x_{\nu}$,
we obtain 
\be
\frac{d\Gamma }{dx_{\gamma}} &=& 
\frac{d\Gamma_{CPC} }{dx_{\gamma}} + 
\frac{d\Gamma_{CPV} }{dx_{\gamma}}\,,
\ee
where
\be
\frac{d\Gamma_{CPC} }{dx_{\gamma}} &=&
{4\alpha\over 3}\left({G_F\alpha\over   
16\pi^2\sin^2\theta_W}\right)^2
\sum_{l=e,\mu,\tau}\left(
Re\lambda_c\;X_{NL}^l+Re\lambda_t\;X(x_t)\right)^2
\nn\\
&&\cdot |F_{V}|^2x_{\gamma}^3(1-x_{\gamma})
M_K^5\,,
\ee
and
\be
\frac{d\Gamma_{CPV} }{dx_{\gamma}} &=&
4\alpha\left({G_F\alpha\over   
16\pi^2\sin^2\theta_W}\right)^2
\left(Im\lambda_t\;X(x_t)F_A\right)^2
x_{\gamma}^3(1-x_{\gamma})
M_K^5\,.
\ee

To illustrate the numerical results, we use $m_d=0.3\ GeV$, $m_s=0.4\
GeV$, $m_{t}=166\ GeV$,
$m_{c}=1.30\ GeV$, $M_{K}=0.5\ GeV$, $\Lambda=325\ MeV$,
$\alpha(M_Z)=1/128$, $\sin^2\theta_W= 0.23$, 
$\omega=0.34$, and the CKM
parameters \cite{review,PDG,CKM-f} of $\lambda=0.22$, $A= 0.83$,
$\rho=0.13$, and $\eta=0.34$.
The differential decay branching ratios of $dB(\knng)_{CPC}/dx_{\gamma}$ 
and $dB(\knng)_{CPV}/dx_{\gamma}$ 
as a function of $x_{\gamma}=2E_{\gamma}/M_K$ are shown in Figs. 1 and 2, 
respectively.
The decay branching ratios are found to be
\be
\label{BrCPC}
B(\knng)_{CPC} &=&1.0\times 10^{-13}\,,
\\
\label{BrCPV}
B(\knng)_{CPV} &=&1.5\times 10^{-15} \,.
\ee
From Eqs. (\ref{BrCPC}) and (\ref{BrCPV}), we find that 
the CP conserving contribution to the decay branching ratio is about
a factor of $67$ larger than that from CP violating one. 
It is clear that the numerical values in (\ref{BrCPC}) and (\ref{BrCPV})
depend on the values of the CKM parameters of $\rho$ and $\eta$,
respectively.
Nevertheless, one could conclude that 
a measurement of the decay would determine the real part of
$V_{td}$.

We now compare our numerical result of the CP conserving contribution in
Eq. (\ref{BrCPC}) with those in Refs. \cite{RP} and \cite{MP}.
Our value is about 
two orders of magnitude 
and a factor 2 
smaller than that in \cite{RP} and \cite{MP}, respectively.
The main reason for the former difference is due to a factor 2 was
missed in Eq. (28) of Ref. \cite{MP}, whereas that for the later one is
unclear. 
It seems that one needs to re-study the approach in
Ref. \cite{RP}.
Finally, we remark that the ratio between the CP conserving and CP
violating branching rates agree with that estimated in Ref. \cite{MP}.

\se{Conclusions}

We have studied the CP conserving and violating contributions 
to the decay of $\knng$ in the standard model. 
With the form factors for $K\to\gamma$ transitions
calculated directly in
the entire physical range of momentum transfer
within the light front framework,  
we have shown that
the CP conserving part is much larger 
than that from CP violating one.
We have found that the decay branching ratio is at the level
of $10^{-13}$,
which could be accessible
at a future kaon project such as the KAMI at FNAL \cite{Hsiung}.

\vspace{2cm}

\noindent
{\bf Acknowledgments}

This work is supported by the National Science Council of the
ROC under contract number NSC89-2112-M-007-013.

\newpage       
\begin{figure}[h]
\includegraphics{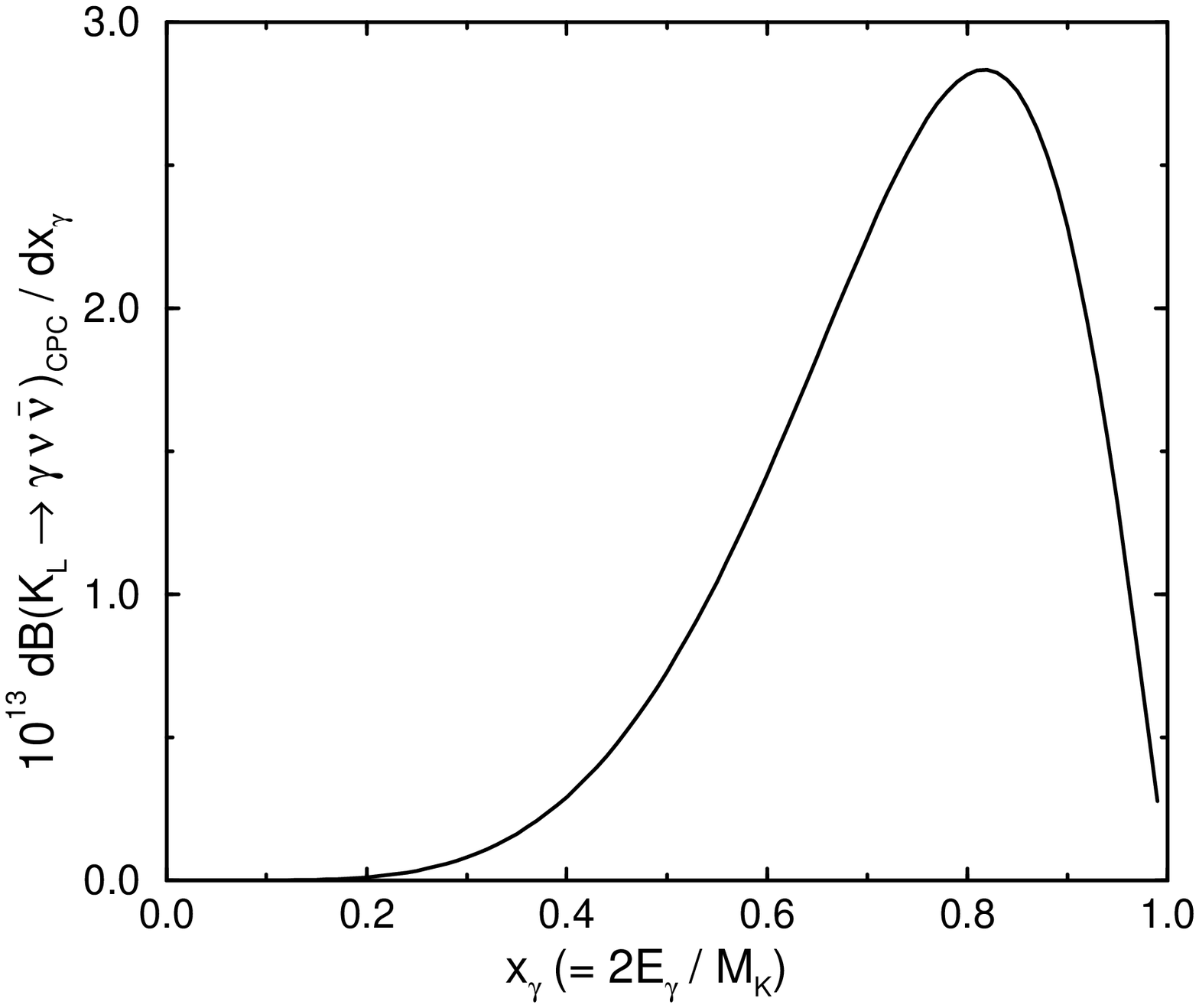}
\end{figure}
\vskip 13.5cm
\bc
Fig. 1.
The differential decay
branching ratios $dB(K_L\to \gamma\nu\bar{\nu})_{CPC}/dx_{\gamma}$ as a
function of $x_{\gamma}=2E_{\gamma}/M_{K}$.
\ec

\newpage       
\begin{figure}[h]
\includegraphics{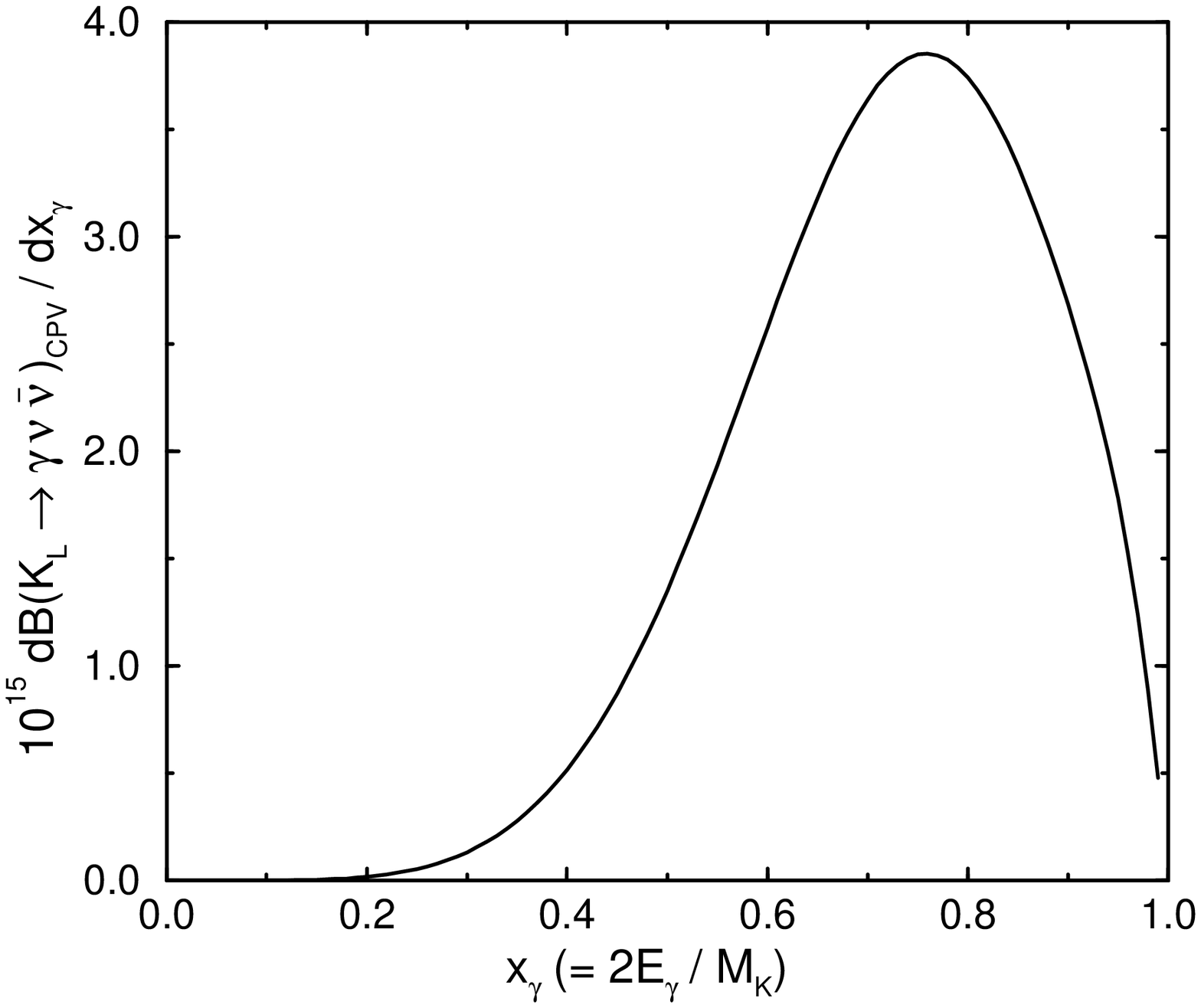}
\end{figure}
\vskip 13.5cm
\bc
Fig. 2.
The differential decay
branching ratios $dB(K_L\to \gamma\nu\bar{\nu})_{CPV}/dx_{\gamma}$ as a
function of $x_{\gamma}=2E_{\gamma}/M_{K}$.
\ec

\ed
\end{document}